\ifx\pdfoutput\undefined
	\documentclass[dvips]{elsart}
\else
	\documentclass[pdftex]{elsart}
\fi

\usepackage{natbib}
\usepackage{graphicx}
\usepackage{amsmath}

\begin{document}

\begin{frontmatter}

\title{Evidence for isotropic emission in GRB991216}

\author[1,2]{R. Ruffini}
\ead{ruffini@icra.it}
\author[1,2]{M.G. Bernardini}
\author[1,2]{C.L. Bianco}
\author[1,3]{P. Chardonnet}
\author[1,4]{F. Fraschetti}
\author[1,2]{S.-S. Xue}

\address[1]{ICRA --- International Center for Relativistic Astrophysics.}
\address[2]{Dip. Fisica, Univ. Roma ``La Sapienza'', P.le A. Moro 5, 00185 Roma, Italy.}
\address[3]{Univ. Savoie, LAPTH - LAPP, BP 110, 74941 Annecy-le-Vieux Cedex, France.}
\address[4]{Univ. Trento, Via Sommarive 14, 38050 Povo (Trento), Italy.}

\begin{abstract}
The issue of the possible presence or absence of jets in GRBs is here re-examined for GRB991216. We compare and contrast our theoretically predicted afterglow luminosity in the $2$--$10$ keV band for spherically symmetric versus jetted emission. At these wavelenghts the jetted emission can be excluded and data analysis confirms spherical symmetry. These theoretical fits are expected to be improved by the forthcoming data of the Swift mission.
\end{abstract}


\end{frontmatter}

\paragraph*{Introduction} We recall that the determination of the presence or absence of jets in Gamma-Ray Burst (GRB) afterglows is examined for GRB 991216 in which a narrow half-opening beaming angle $\vartheta_\circ=3^\circ$ has been claimed \citep{ha00}. We show that the afterglow is consistent with a spherically symmetric distribution. The choice of GRB 991216 is motivated by the superb set of data existing in the ``prompt'' emission (in the band $50$--$300$ keV observed by \citealp{brbr99}) and in the afterglow (in the band $2$--$10$ keV observed by R-XTE and Chandra, see \citealp{cs00,ha00,p00}). In recent publications \citep[see][and references therein]{Brasile} we have developed basic formulas for the GRB model. This development differs somewhat from the usual presentation in the literature. The differences includes:
\textbf{a)}
the entire spacetime parametrization of the GRB phenomenon starting from the moment of gravitational collapse, to the optically thick accelerated phase, all the way to the afterglow \citep{lett1};
\textbf{b)}
the identification of the ``prompt'' radiation as the early emission in the afterglow era \citep{lett2,rbcfx02_letter};
and \textbf{c)}
a marked distinction between the sharp X and $\gamma$ radiation, which is energetically predominant in the afterglow ($\sim 90$\%), and the highly variable radiation in the optical and radio bands which is generally much weaker, observed in the late afterglow phases and, in some cases, absent \citep{Spectr1}.
The usual presentations in the literature consider the afterglow radiation in all wavelength from the $\gamma$ rays all the way to the radio as originating from a unique physical process mainly related to synchrotron radiation \citep[see e.g.][and references therein]{p04}. In our approach we assume instead that the X and $\gamma$ radiation originates from the emission in the shock front with a thermal spectrum in the co-moving frame \citep{Spectr1}. The optical and radio emission is assumed to be emitted by the matter compressed in the pre-shock region, possibly with contributions from magnetic fields and synchrotron emission \citep{Spectr1,Spectr2}. This progress has allowed to obtain very specific theoretical predictions for the luminosity of the entire afterglow in fixed X and $\gamma$ energy bands \citep{Spectr1,Spectr2}. We have fitted by a unified theoretical approach the ``prompt'' radiation and the decaying part of the afterglow. The ``prompt'' radiation has been shown to coincide with the emission of the peak of the afterglow \citep{lett1,lett2,rbcfx02_letter}.

\paragraph*{The free parameters and the equations of our model and the fit of GRB 991216} Our model depends: \textbf{a)} on the initial conditions at the beginning of the afterglow era, which are functions of the only two free parameters describing the source: the total energy $E_{tot}$, which coincides with the dyadosphere energy $E_{dya}$ \citep[see][and references therein]{Brasile}, and the amount $M_B$ of baryonic matter left over from the gravitational collapse of the progenitor star, which is determined by the dimensionless parameter $B=M_Bc^2/E_{dya}$ \citep[see][]{rswx00}; \textbf{b)} on the equations of motion of the accelerated baryonic matter pulse which, interacting with the ISM, gives origin to the afterglow \citep[][and references therein]{Brasile}; and \textbf{c)} on the afterglow EQuiTemporal Surfaces \citep[EQTS, see][]{rbcfx02_letter,EQTS_ApJL,EQTS_ApJL2} as well as on the two independent variables describing the interstellar medium (ISM): the ISM density $n_{ism}$ and the parameter $\mathcal{R}={A_{eff}}/{A_{abm}}$, which gives the ratio between the ``effective emitting area'' $A_{eff}$ and the accelerated baryonic matter (ABM) pulse surface area $A_{abm}$. This factor $\mathcal{R}$, with the density $n_{ism}$, is sufficient to identify the ISM filamentary structure and the basic physical process originating the X and $\gamma$ flux in GRB afterglows \citep{Spectr1,Spectr2}. For GRB 991216 we have \citep[see][]{Brasile}: $E_{dya} = 4.83 \times 10^{53}$ erg, $B = 2.7 \times 10^{-3}$. This leads to the following conditions at the beginning of the afterglow era for the photon detector arrival time, the ABM pulse radius, its Lorentz gamma factor and its baryonic matter content respectively: $t_a^d = 6.11 \times 10^{-2}$ s, $r_\circ = 1.61 \times 10^{14}$ cm, $\gamma_\circ = 340.3$, $M_B = 1.45 \times 10^{30}$ g. Starting from the above initial conditions, we have obtained the equations of motion for the baryonic matter giving rise to the afterglow radiation as well as the precise expressions of the EQTSs shown in \citet{EQTS_ApJL,EQTS_ApJL2}.

The temperature $T$ of the black body in the comoving frame is:
\begin{equation}
T=\left(\frac{dE_\mathrm{int}}{4\pi r^2 d\tau \sigma \mathcal{R}}\right)^{1/4}\, ,
\label{tcom}
\end{equation}
where $\sigma$ is the Stefan-Boltzmann constant and $dE_\mathrm{int}$ is the energy developed during the co-moving time $d\tau$ by the collision \citep[see][]{Spectr1}. The ABM pulse is formed by en electrons-nucleons plasma and has an intrinsic ``rigidity'' due to its ultrarelativistic motion which amplifies the tangential component of the electromagnetic field. The ABM pulse, therefore, responds coherently as a whole to the collision with the ISM \citep[details in][and references therein]{Brasile}. The source luminosity at a detector arrival time $t_a^d$, per unit solid angle $d\Omega$ and in the energy band $\left[\nu_1,\nu_2\right]$ is given by \citep[see][]{Brasile,Spectr1}:
\begin{equation}
\frac{dE_\gamma^{\left[\nu_1,\nu_2\right]}}{dt_a^d d \Omega } = \frac{1}{4\pi}\int_{EQTS} \left[d \varepsilon \frac{v \cos \vartheta}{\Lambda^{4}} \frac{dt}{dt_a^d} W\left(\nu_1,\nu_2,T_{arr}\right)\right] d \Sigma\, ,
\label{fluxarrnu}
\end{equation}
where $d \varepsilon=d E_{int}/V$ is the energy density released in the interaction of the ABM pulse with the ISM inhomogeneities measured in the comoving frame, $\vartheta$ is the angle between the radial expansion velocity of a point on the pulse surface and the line of sight, $\Lambda=\gamma(1-(v/c)\cos\vartheta)$ is the Doppler factor, $W\left(\nu_1,\nu_2,T_{arr}\right)$ is an ``effective weight'' required to evaluate only the contributions in the energy band $\left[\nu_1,\nu_2\right]$, $d\Sigma$ is the surface element of the EQTS at detector arrival time $t_a^d$ on which the integration is performed \citep[see][]{rbcfx02_letter,EQTS_ApJL,EQTS_ApJL2} and $T_{arr}$ is the observed temperature of the radiation emitted from $d\Sigma$:
\begin{equation}
T_{arr}=\frac{T}{\gamma \left(1-\left(v/c\right)cos\vartheta\right)}\frac{1}{(1+z)}\, .
\label{Tarr}
\end{equation}
The thermalization process due to the optically thick condition generated by the ISM filamentary structure has been discussed in \citet{Spectr2}. The ``effective weight'' $W\left(\nu_1,\nu_2,T_{arr}\right)$ is given by the ratio of the integral over the given energy band of a Planckian distribution at a temperature $T_{arr}$ to the total integral $aT_{arr}^4$:
\begin{equation}
W\left(\nu_1,\nu_2,T_{arr}\right)=\frac{1}{aT_{arr}^4}\int_{\nu_1}^{\nu_2}\rho\left(T_{arr},\nu\right)d\left(\frac{h\nu}{c}\right)^3\, ,
\label{effweig}
\end{equation}
where $\rho\left(T_{arr},\nu\right)$ is the Planckian distribution at temperature $T_{arr}$:
\begin{equation}
\rho\left(T_{arr},\nu\right)=\frac{2}{h^3}\frac{h\nu}{e^{h\nu/\left(kT_{arr}\right)}-1}
\, .
\label{rhodef}
\end{equation}
\begin{figure}
\centering
\includegraphics[width=6.8cm,clip]{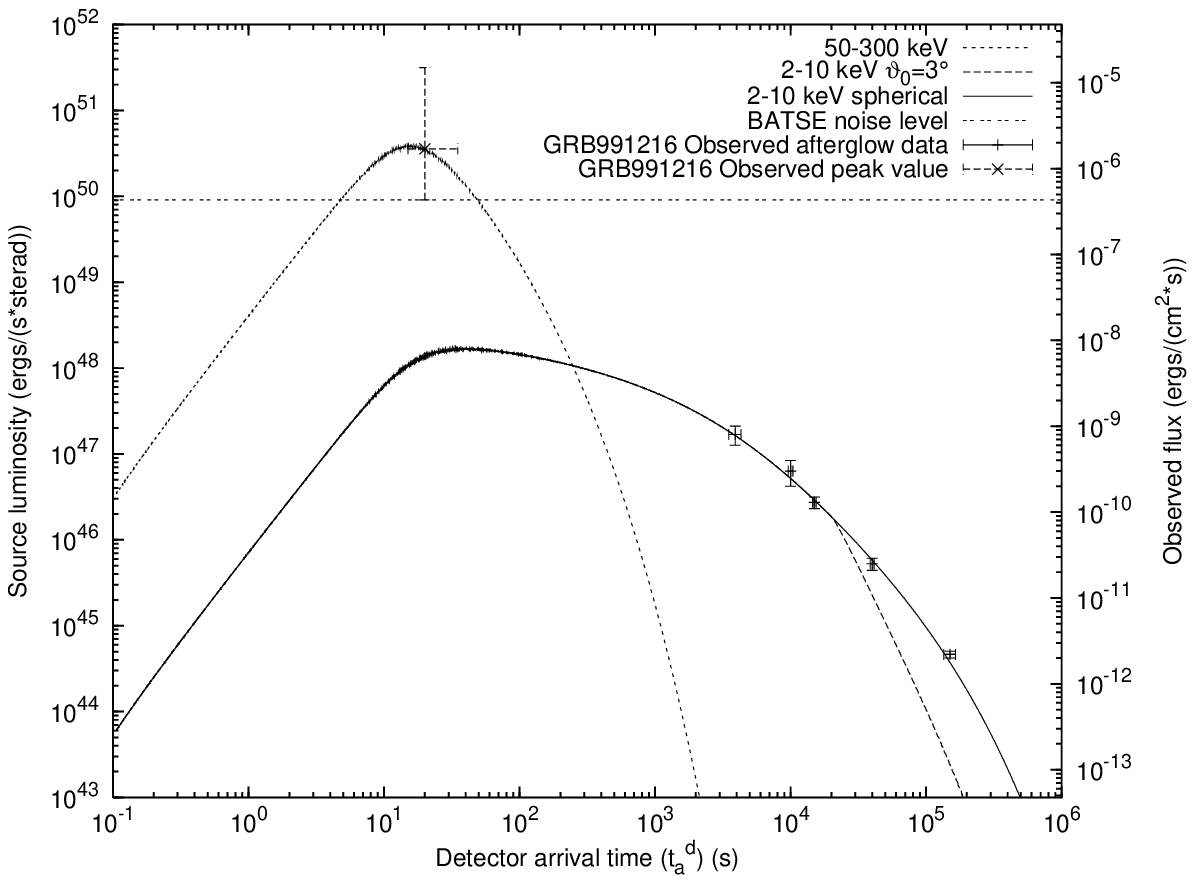}
\includegraphics[width=6.8cm,clip]{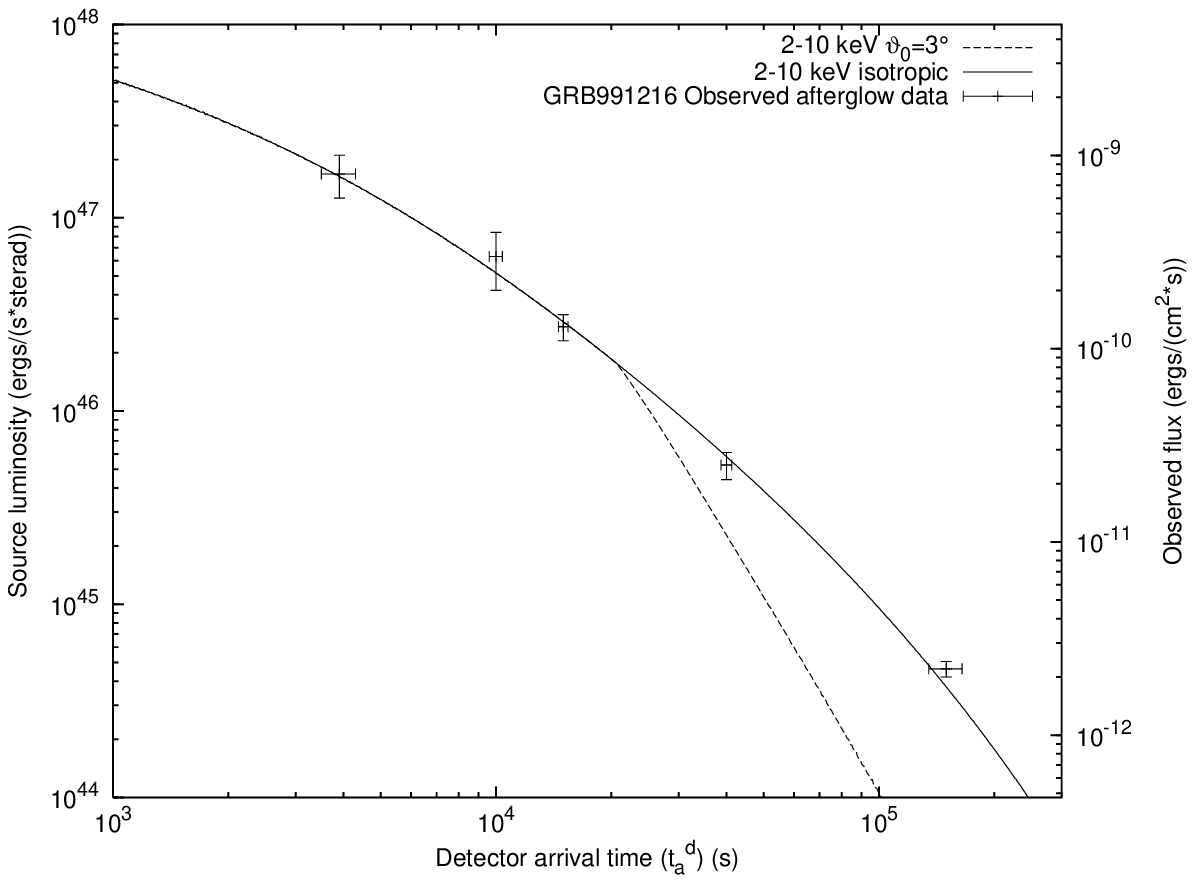}
\caption{\textbf{Left panel:} Best fit of the afterglow data of GRB991216. The dotted curve  is the luminosity in the $50$--$300$ keV energy band. The solid curve is the luminosity in the $2$--$10$ keV band computed assuming spherical symmetry. The observational data from R-XTE and Chandra \citep[see][]{ha00} are perfectly consistent with such an assumption. The presence of a $\vartheta_\circ=3^\circ$ half-opening beaming angle (dashed curve) is ruled out. \textbf{Right panel:} Enlargement of the plot in the region of the afterglow observational data from R-XTE and Chandra.}
\label{991216}
\end{figure}

The estimate of the theoretically predicted luminosity in a fixed energy band as a function of the initial data is then perfectly well defined both for the ``prompt'' radiation and the decaying part of the afterglow. Almost $10^8$ paths with different temperatures and different Lorentz boosts have to be considered in the integration over the EQTSs. We give in Fig. \ref{991216} the results in the two energy bands $50$--$300$ keV (observed by BATSE) and $2$--$10$ keV (observed by R-XTE and Chandra). It is most remarkable that the best fit of the descending part of the afterglow (for a photon arrival time at the detector $t_a^d > 10^3$ s) is obtained by a constant factor $\mathcal{R} = 1.0 \times 10^{-10}$ and a constant $n_{ism} = 3.0$ particles/cm$^3$. We point out the agreement with the data of the ``prompt'' radiation obtained by BATSE in the energy range $50$--$300$ keV (see the dotted line in Fig. \ref{991216}). We have succeeded as well in the fit of the data obtained by the R-XTE and Chandra satellites \citep{ha00} in the energy range $2$--$10$ keV (see dashed line in Fig. \ref{991216}). These data refer to the decaying part of the afterglow. These fits cover a time span of $\sim 10^6$ s and it is remarkable that they are a sole function of the two variables $\mathcal{R}$ and $n_{ism}$ which have a constant value in this region. We have also computed, within this global self-consistent approach which fits both the ``prompt'' radiation and the decaying part of the afterglow, the flux in the $2$--$10$ keV range which would be expected for a beamed emission with half opening angle $\vartheta_\circ = 3^\circ$, see Fig. \ref{991216}. The presence of beaming manifest itself, as expected, in the decaying part of the afterglow and is incompatible with the data.

\paragraph*{Conclusions} Our theory sharply differs from the so-called ``state of the art'' in this field. Instead of the traditional multiwavelenght approach, we differentiate the mechanisms for the sharp X and $\gamma$ radiation, originating in the shock front with a thermal spectrum in the co-moving frame, from the ones originating the radio and optical radiation, which we expect to be emitted ahead of the shock front by more traditional processes. We expect improvements in the accuracy of the fits by the advent of the Swift satellite. We also like to point out that additional and independent evidence for GRB spherical symmetry comes from the fit of the spectral data \citep{Spectr1}. We can then draw the following general conclusions: \textbf{1)} It is clear that a spherically symmetric expansion of the GRB afterglow is perfectly consistent with the data, rather than a narrow jet as previous authors have concluded. \textbf{2)} The actual afterglow luminosity in fixed energy bands, in spherical symmetry, does not have a simple power law dependence on arrival time (see Fig. \ref{991216}). This circumstance has been erroneously interpreted, in the usual presentation in the literature, as a broken power-law supporting the existence of jet-like structures in GRBs. Moreover, the slope of the beamed emission and the arrival time at which the break occurs have been there computed using the incorrect equations of motion for the afterglow and the incorrect EQTSs \citep{EQTS_ApJL,EQTS_ApJL2}. \textbf{3)} If one assumes the presence of jets in a consistent afterglow theory, one finds that the break corresponding to the purported beaming appears at an arrival time incompatible with the observations (see right panel in Fig. \ref{991216}). In addition to the source GRB991216, our model has been applied successfully, assuming spherical symmetry, to GRB980425 \citep[see][]{cospar02}, GRB030329 \citep[see][]{r03mg10} and to GRB970228 \citep[see][]{cgrb03}. The GRB spherical symmetry unambiguously points to the electromagnetic energy component of the black hole extractable energy as the GRB energy source \citep{cr71,dr75}.

\end{document}